\documentclass[10pt, prd, twocolumn, showpacs, amsmath, amssymb, aps, floats, floatfix, nofootinbib]{revtex4-1}

\usepackage{bm}
\usepackage{array}
\usepackage{graphicx}
\usepackage{subfigure}
\usepackage{multirow}
\usepackage{dcolumn}
\usepackage[]{color}
\usepackage[CJKbookmarks=true,colorlinks,linkcolor=blue,anchorcolor=red,citecolor=green]{hyperref}
\usepackage{cleveref}
\usepackage{multirow}
\usepackage{setspace}

\def\kpc{\mathrm{kpc}}
\def\km{\mathrm{km}}

\def\GeV{\mathrm{GeV}}

\def\cm{\mathrm{cm}}
\def\s{\mathrm{s}}

\newcolumntype{p}{D{,}{\pm}{-1}}

\begin{document}
\title{Determination of dark matter distribution in Ursa Major III and constraints on dark matter annihilation}

\author{Yi Zhao$^{1}$}
\author{Xiao-Jun Bi$^{2,3}$}
\author{Su-Jie Lin$^{4}$}
\author{Peng-Fei Yin$^{2}$}

\affiliation{
$^1$ College of Physics and Materials Science, Tianjin Normal University, Tianjin 300387, China \\
$^2$ Key Laboratory of Particle Astrophysics, Institute of High Energy Physics, Chinese Academy of Sciences, Beijing 100049, China \\
$^3$ School of Physical Sciences, University of Chinese Academy of Sciences, Beijing 100049, China \\
$^4$ School of Physics and Astronomy, Sun Yat-Sen University, Zhuhai, Guangdong, 519082, China
}

\begin{abstract}
The recently discovered satellite dwarf galaxy Ursa Major III provides a promising opportunity to explore the signatures resulting from dark matter (DM) annihilation, due to its proximity and large J-factor. Owing to the absence of an excess of $\gamma$-ray signatures originating from Ursa Major III, observations of $\gamma$-rays, such as those from  Fermi-LAT, can be utilized to set constraints on the DM annihilation cross section.
In this study, we determine the DM density profile, and consider the relationship between DM density and velocity dispersion at different locations within Ursa Major III through Jeans analysis.
We calculate the J-factor of Ursa Major III for s-wave annihilation, along with the effective J-factors for p-wave and Sommerfeld enhanced annihilation scenarios.
Utilizing these derived J-factors, we set stringent constraints on DM annihilation cross sections in three scenarios.
Given the substantial impact of member star identification on the J-factor of Ursa Major III, we further calculate J-factors with the condition of excluding the largest velocity outlier. Our analysis reveals a notable reduction in the median value and an increase in the deviation of J-factors, thereby leading to considerably weaker constraints.

\end{abstract}


\maketitle

\section{Introduction}

One popular candidate for cold dark matter (DM) is the weakly interacting massive particles (WIMPs) ~\cite{Jungman:1995df,Bergstrom:2000pn,Bertone:2004pz}. The current abundance of WIMP can explain the observed DM relic density in the thermal freeze-out scenario. The annihilation of WIMPs can directly or indirectly produce $\gamma$-rays. These $\gamma$-rays are expected to be predominantly generated in regions with high DM densities and can be detected by satellite and terrestrial experiments.

Dwarf spheroidal galaxies (dSphs) have long been regarded as promising targets for detecting such signatures. This is attributed to their proximity, high DM densities, and the absence of conventional astrophysical $\gamma$-ray sources~\cite{Mateo:1998wg,Grcevich:2009gt}. These characteristics make dSphs ideal candidates for probing DM signatures.
While significant signatures have not yet been detected, studies in the literature have provided valuable constraints on the DM annihilation cross section~\cite{Fermi-LAT:2010cni,Fermi-LAT:2011vow,Fermi-LAT:2016uux,Geringer-Sameth:2011wse,Cholis:2012am,Geringer-Sameth:2012lsj,
Mazziotta:2012ux,Baushev:2012ke,Huang:2012yf,Fermi-LAT:2013sme,Fermi-LAT:2015att,Tsai:2012cs,Essig:2010em,Choquette:2016xsw,Lu:2017jrh,Strigari:2018utn,Hoof:2018hyn,Hess:2021cdp, DiMauro:2022hue,McDaniel:2023bju,Boddy:2024tiu,LHAASO:2024upb}.
These constraints primarily rely on the profile and proximity of the discussed dwarf spheroidal halos, characterized by an integrated parameter known as the J-factor.
Some dSphs exhibit high DM density and close proximity, with J-factors that notably surpass others, therefore playing a crucial role in shaping the constraint.
The recently identified Ursa Major III/UNIONS 1, located at a heliocentric distance of $\sim 10\kpc$ and unveiled by the Ultraviolet Near Infrared Optical Northern Survey ~\cite{Errani:2023sgd,smith2023discovery}, emerges as a potential candidate for such a dSph.

Ursa Major III/UNIONS 1 contains an old and metal-poor stellar population.
If this system is a self-gravitating star cluster, the Galactic tidal field would disperse it within approximately 0.4 $\mathrm{Gyr}$.
The detection of Ursa Major III/UNIONS 1 in this scenario can be viewed as an accidental event, providing a unique window into its final orbital path around the Milky Way.
If this observation is not a mere coincidence, it would necessitate the presence of a DM halo with a mass of $\sim10^9M_\odot$, thus rendering the system a remarkably close dSph~\cite{Errani:2023sgd}.
In the context of this study, we refer to this system as the dSph Ursa Major III hereafter.

Despite the relatively lower mass of its DM halo in comparison to other massive dSphs, the close proximity of Ursa Major III makes it a notable source of gamma-ray emissions resulting from DM annihilation, boasting a potentially large J-factor.
Through the application of the analytic formula incorporating the heliocentric distance, projected half-light radius, and the line-of-sight velocity dispersion of the dSph~\cite{Evans:2016xwx,Pace:2018tin}, the J-factor within $0.5^\circ$ of the Ursa Major III center is estimated to be on the order of $\sim 10^{21}\GeV^2\s^{-5}$ ~\cite{Errani:2023sgd,Crnogorcevic:2023ijs}.
For comparison, the J-factors of conventional dSphs rarely exceed $10^{20}\GeV^2\s^{-5}$.

The remarkable J-factor of Ursa Major III highlights its potential to impact the constraints on DM annihilation significantly.
Utilizing 15 years of Fermi-LAT data from Ursa Major III, the authors of Ref.~\cite{Crnogorcevic:2023ijs} have established notably more stringent constraints on velocity-independent DM annihilation cross sections, compared to previous constraints derived from observations of other dSphs.
As DM particles exhibit different velocity dispersions in diverse astrophysical systems, their annihilation cross sections may also manifest significant variations in velocity-dependent annihilation scenarios.
Consequently, the J-factors and corresponding constraints on the local DM annihilation cross section derived from observations of dSphs necessitate adjustments \cite{Zhao:2016xie,Robertson:2009bh,Boddy:2017vpe,Zhao:2017dln,Bergstrom:2017ptx,Petac:2018gue,Johnson:2019hsm,Boddy:2019wfg,Boddy:2019qak,Board:2021bwj,Boucher:2021mii,Kiriu:2022bjq,Blanchette:2022hir}.
In this context, in addition to investigating the conventional velocity-independent s-wave annihilation scenario, we explore the implications of velocity-dependent annihilation processes, by incorporating the observations of Ursa Major III. The corresponding effective J-factors for velocity-dependent annihilation should encompass information about the DM velocity distribution \cite{Robertson:2009bh,Boddy:2017vpe,Zhao:2017dln,Bergstrom:2017ptx,Petac:2018gue,Johnson:2019hsm,Boddy:2019wfg,Boddy:2019qak,Board:2021bwj,Boucher:2021mii,Kiriu:2022bjq,Blanchette:2022hir}.
To elucidate the relationship between the DM annihilation cross section and velocity dispersion at different spatial positions, we can determine the DM density profile and resolve the DM velocity dispersion through the Jeans analysis~\cite{Zhao:2017dln}.

In this study, we employ the Markov Chain Monte Carlo (MCMC) toolkit GreAT, which is integrated within the CLUMPY package~\cite{Bonnivard:2015pia}, to conduct a Jeans analysis.
This framework enables us to derive the DM density profile of Ursa Major III, by incorporating the current results from stellar kinematic observations.
Based on the DM profiles derived from the Jeans analysis, we determine the
J-factor for s-wave annihilation, alongside the effective J-factors for p-wave annihilation and Sommerfeld-enhanced annihilation in the Coulomb limit.
The effective J-factors derived from this analysis have important applications in specific scenarios, allowing us to establish constraints on velocity-dependent DM annihilation.

It is worth noting that the aforementioned large J-factor of Ursa Major III is estimated through an analysis encompassing 11 member stars.
Upon the exclusion of the largest velocity outlier, the intrinsic line-of-sight velocity dispersion experiences a notable reduction from $3.7^{+1.4}_{-1.0} \km/\s$ to $1.9^{+1.4}_{-1.1}\km/\s$ \cite{smith2023discovery}. Consequently, a wide J-factor range spanning approximately $10^{19} - 10^{22} \GeV^2\cm^{-5}$ is derived in Ref.~\cite{Errani:2023sgd}.
To provide a comprehensive analysis, we further determine the DM profiles and J-factors through the Jeans analysis under the excluding of the largest velocity outlier.
This comparative approach shows the influence of the identification of member stars on the DM profile and associated J-factor calculations for Ursa Major III.

This paper is organized as follows. In Sec. 2, we introduce the Jeans analysis methodology employed to ascertain the DM density profile of Ursa Major III. In Sec. 3, we present the J-factor for velocity-independent annihilation and the effective J-factors for two velocity-dependent annihilation processes. We set constraints on DM annihilation cross sections based on results from Fermi-LAT. In Sec. 4, we calculate the J-factors under the condition of excluding the largest velocity outlier. Finally, Sec. 5 is the discussion and conclusion.

\section{Jeans Analysis}
\label{sec_2}

In this study, we conduct a Jeans analysis to derive the DM density profile of Ursa Major III. The dynamics of a stellar system under the influence of a gravitational field are governed by the Jeans equation, which is derived from the collisionless Boltzmann equation. Assuming spherical symmetry, a steady-state system, and negligible rotational support, the second-order Jeans equation is simplified to ~\cite{Courteau:2013cjm}
\begin{eqnarray}\label{eq:zy1}
\frac{1}{\nu(r)}\frac{d}{dr}[\nu(r)\sigma_r^2]+2\frac{\beta_{ani}(r)\sigma_r^2}{r}=-\frac{GM(r)}{r^2},
\end{eqnarray}
where $G$ denotes the gravitational constant, $\nu(r)$ represents the three-dimensional stellar number density, $\sigma_r^2$ is the radial velocity dispersion of stars in dSphs, $M(r)$ is the enclose mass given by $M(r)=4\pi\int_0^r\rho_{DM}(s)s^2ds$, and $\beta_{ani}(r)=1-\sigma_\theta^2/\sigma_r^2$ denotes the stellar velocity anisotropy, which depends on the ratio of tangential to radial velocity dispersions. The utilization of the DM mass density $\rho_{DM}$ in the enclose mass, instead of the total mass density profile, is motivated by the minor contribution of the stellar component in comparison to the DM halo.

The solution to the above Jeans equation can be expressed as
\begin{eqnarray}\label{eq:zy2}
\nu(r)\sigma_r^2=\frac{1}{A(r)}\int_r^{\infty}A(s)\nu(s)\frac{GM(s)}{s^2}\mathrm{d}s,
\end{eqnarray}
where $A(r) \equiv A_{r_1}\exp[\int_{r_1}^r\frac{2}{t}\beta_{ani}(t)dt]$. Here the mute parameter $r_1$
only results in a normalization factor that cancels out in Eq. \ref{eq:zy2} ~\cite{Bonnivard:2014kza}. Note that only the two-dimensional projected stellar number density and the line-of-sight velocity dispersion commonly provided by astrophysical observations. Consequently, the Jeans equation must be adapted to account for these two-dimensional observations, leading to a solution given by
\begin{eqnarray}\label{eq:zy3}
\sigma_p^2(R)=\frac{2}{I(R)}\int_R^{\infty}[1-\beta_{ani}(r)\frac{R^2}{r^2}]\frac{\nu(r)\sigma_r^2r}{\sqrt{r^2-R^2}}\mathrm{d}r,
\end{eqnarray}
where $I(R)$ represents the projected light profile, specifically surface brightness, and
$\sigma_p(R)$ denotes the projected velocity dispersion corresponding to the projected radius $R$.

In our study, we adopt the NFW profile~\cite{Navarro:1996gj} to model the DM halo, which is expressed as
\begin{eqnarray}\label{eq:zy8}
\rho_{DM}=\frac{\rho_s}{\dfrac{r}{r_s}\left(1+\dfrac{r}{r_s}\right)^2},
\end{eqnarray}
where the normalization $\rho_s$ and scale radius $r_s$ are treated as free parameters in the MCMC analysis. Furthermore, the anisotropy parameter $\beta_{ani}$ in Eq.~\ref{eq:zy3} is also taken as a free parameter. In order to derive $\sigma_p(R)$ from Eq.~\ref{eq:zy3}, the determination of $I(R)$ is necessary.

The authors in Ref.~\cite{smith2023discovery} estimated the surface brightness of Ursa Major III by assuming that its member stars are distributed according to an elliptical and exponential radial surface density profile, including constant field contamination.
The profile is given by
\begin{eqnarray}\label{eq:zy5}
\rho_{\mathrm{stars}}=N\frac{1.68^2}{2\pi r_h^2 (1-\epsilon)}\exp\left(\frac{-1.68r}{r_h}\right),
\end{eqnarray}
with the elliptical radius
\begin{eqnarray}\label{eq:zy6}
r &=& {\bigg\{\frac{1}{(1-\epsilon)^2}\left[(x-x_0)\cos\theta-(y-y_0)\sin\theta \right]^2 \nonumber \bigg.}\\
&+& {\bigg. \left[(x-x_0)\sin\theta+(y-y_0)\cos\theta \right]^2\bigg\}^{\frac{1}{2}} },
\end{eqnarray}
where $\epsilon$ represents the ellipticity, $\theta$ is the position angle of the major axis, the half-light radius $r_h$ corresponds to the length of the semi-major axis, $N$ is the number of the stars in the system, $(x_0,y_0)$ denotes the central coordinates of the profile.
The authors have provided surface brightness by giving the median values of these parameters along with their 1$\sigma$ deviations. Using these results, we can estimate the averaged surface brightness at various radii from the galaxy center.

\begin{figure}[!htb]
\centering
\includegraphics[width=0.9\columnwidth, angle=0]{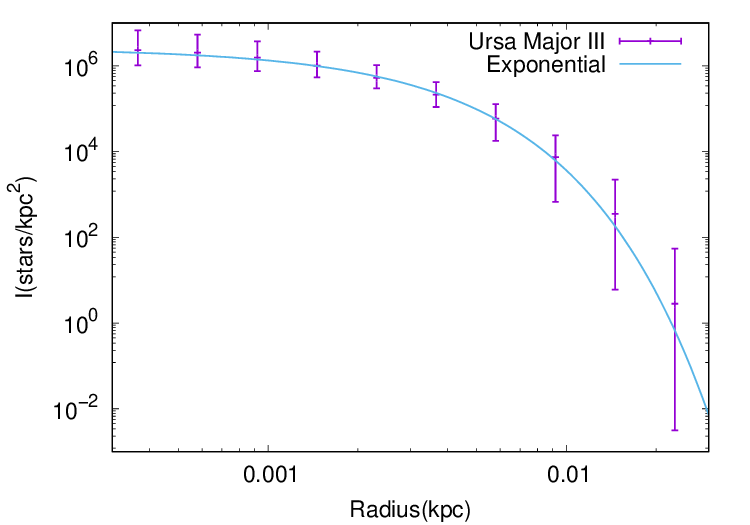}
\caption{The stellar density of Ursa Major III. The blue solid line
represents the best fit given by the exponential model.}
\label{fig:zy1}
\end{figure}

The projected light profile $I(R)$ can be obtained by fitting this surface brightness to an exponential model~\cite{Evans:2008ik}
\begin{eqnarray}\label{eq:zy7}
I(R)=I_0 \exp(-\frac{R}{r_c}).
\end{eqnarray}
The best fit is achieved with $I_0=2.58\times 10^6\,\mathrm{stars}/\kpc^{2}$ and $r_c=1.52\times 10^{-3}\,\kpc$.
We illustrate this best-fit profile in comparison with the averaged surface brightness derived from the results given by Ref.~\cite{smith2023discovery} in Fig. \ref{fig:zy1}.

The kinematic data required for the Jeans analysis include the coordinates and line-of-sight velocities of the stars, along with their corresponding errors for each individual.
In classical dSphs with substantial stellar populations, a considerable proportion of stars exhibit in their membership status, neither definitively belonging nor excluded from the galaxy. Consequently, incorporating membership probabilities for each star is crucial.
For ultrafaint dSphs, distinguished by a limited number of member stars, observations often provide binary classifications ($P_i=0$ or $P_i=1$) for individual stars.
Ursa Major III is such an ultrafaint dSph containing only 11 identified member stars with velocity measurements.
It is worth noting that excluding the largest velocity outlier from the 11 members, despite having a membership probability of 1, leads to a substantial reduction in the intrinsic velocity dispersion~\cite{smith2023discovery}.
In our Jeans analysis, these 11 member stars serve as the primary dataset for determining the DM density profile of Ursa Major III.
Subsequently, we will also discuss the implications of excluding the aforementioned velocity outlier star in a subsequent Sec.~\ref{Sec.IV}.

With given parameters $r_s$, $\rho_s$, and $\beta_{ani}$, the projected velocity dispersion $\sigma_p(R)$ of an individual star at the projected radius $R$ can be calculated by Eq.~\ref{eq:zy3}. The unbinned likelihood function is expressed as
\begin{eqnarray}\label{eq:zy4}
\mathcal{L}_{\mathrm{unbin}}=\prod_{i=1}^{N_{\mathrm{stars}}}\left(\frac{\exp\left(-\dfrac{1}{2}\dfrac{(v_i-\bar{v})^2}{\sigma_p^2(R_i)+\Delta_{v_i}^2}\right)}{\sqrt{2\pi\left[\sigma_p^2(R_i)+\Delta_{v_i}^2\right]}}\right)^{P_i},
\end{eqnarray}
where $v_i$ represents the velocity of the $i$-th star, $\Delta_{v_i}$ denotes the uncertainty in the velocity measurement, and $P_i$ indicates the membership probability for each star. The line-of-sight velocity is assumed to follow a Gaussian distribution, with $\bar{v}$ denoting the mean velocity of the distribution. The observational data for $R_i$, $v_i$, and $\Delta_{v_i}$ of individual stars are taken from Tab. 3 of Ref.~\cite{smith2023discovery}.

The MCMC method is utilized to infer the posterior probability distributions of the parameters from the observational data. Following the principles of Bayes' theorem, the posterior probability of a set of model parameters $\vec{\theta}$ given the data $D$ is denoted as $\mathcal{P}(\vec{\theta}|D) \propto \mathcal{P}(D|\vec{\theta})\mathcal{P}(\vec{\theta})$, where $\mathcal{P}(D|\vec{\theta})= \mathcal{L}(\vec{\theta})$ represents the likelihood function, and $\mathcal{P}(\vec{\theta})$ is the prior probability of the model parameters. We take uniform prior probabilities for the parameters $\log_{10}(\rho_s)$, $\log_{10}(r_s)$, and $\beta_{ani}$. The prior ranges for $\rho_s$, $r_s$, and $\beta_{ani}$ are $[10^5, 10^{13}] \rm M_{\odot}/\rm kpc^3$, $[10^{-2}, 10] \rm kpc$, and $[-9, 1]$, respectively.

To solve the spherical Jeans equation and perform a MCMC analysis, we utilize the MCMC toolkit GreAT, integrated within the CLUMPY package ~\cite{Bonnivard:2015pia}. When dealing with kinematic data presented as line-of-sight velocities for individual stars, the unbinned likelihood method within the CLUMPY package is a better choice. This method offers a distinct advantage over the binned likelihood approach by mitigating uncertainties associated with both the observed velocity dispersion and radius.

We employ the Metropolis-Hastings algorithm~\cite{Hastings:1970aa} to generate eight independent MCMC chains, each consisting of $10^4$ iterations. More than $6000$ DM density profiles with parameters following the posterior probability distribution are obtained. For each DM density profile, we are able to calculate the J-factors for Ursa Major III, including both the traditional J-factor for velocity-independent DM annihilation and the effective J-factors for velocity-dependent annihilation scenarios. Utilizing the derived DM density profiles, we calculate the median values along with their associated deviations for the J-factors.
These values are utilized to establish constraints on the DM annihilation cross section based on $\gamma$-ray observations in the subsequent section. 

In various DM indirect search methodologies, the DM source is often treated as a point source. In such instances, the parameters of the density profile may exhibit degeneracy within the source term, leading to their uncertainties being reflected in the uncertainty of the J factor. Alternatively, if the DM source is extended or the positional information cannot be easily integrated out in the source term, one can directly utilize the derived DM densities profiles to set constraints on the DM properties.

\begin{figure}[!htb]
\centering
\includegraphics[width=0.9\columnwidth, angle=0]{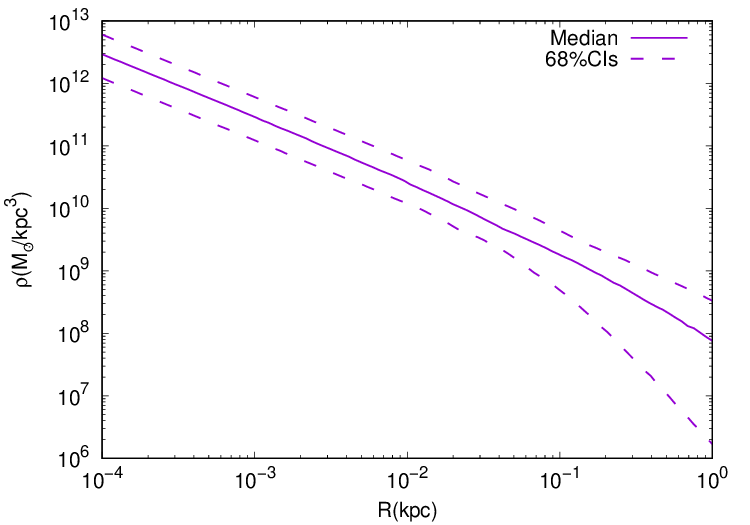}
\caption{The statistical results of the DM density within Ursa Major III. The solid line represents the median value, while the dashed lines correspond to the $68\%$ CIs.}
\label{fig:zy2}
\end{figure}

For the purpose of illustration, we present the statistical results of the DM density at various radial distances from the center in Figure \ref{fig:zy2}. Given a specific radius $R$, we are able to compute the DM densities based on the derived DM density profiles, and determine the median value along with its corresponding deviation.
The solid line denotes the median value of the DM density at the specific radius, while the dashed lines illustrate the $68\%$ confidence intervals (CIs). 
Notably, for DM densities at small radii (below 0.1 kpc), the upper and lower bounds of the $68\%$ CI band are approximately double and half the median values, respectively.

\section{Effective J-factors and Constraints on DM annihilaiton}
\label{sec_3}

We assume that the DM annihilation cross section can be modeled as
\begin{eqnarray}\label{eq:zy9}
\sigma v_{\mathrm{rel}} = a \cdot F(v_{\mathrm{rel}}) \equiv a \cdot (v_{\mathrm{rel}}/c)^n,
\end{eqnarray}
where $v_{\mathrm{rel}}= |\textbf{v}_1- \textbf{v}_2|$ represents the relative velocity of two annihilating DM particles.
In our analysis, we consider three specific values of $n$: (i) $n=0$, corresponding to s-wave velocity-independent annihilation, (ii) $n=2$, corresponding to p-wave annihilation, and (iii) $n=-1$, corresponding to Sommerfeld-enhanced annihilation in the Coulomb limit ~\cite{Sommerfeld:1931qaf} \footnote{The full Sommerfeld enhancement exhibits a complex form that depends on the DM velocity, the coupling of the interaction, and the masses of the DM and mediator. Specifically, only in scenarios with very low DM velocity and a small mass ratio between the mediator and DM, the Sommerfeld enhancement would take the form of $\sim v^{-1}$. For more discussions on the Sommerfeld enhancement in dSphs, refer to, for instance, Ref.~\cite{Ando:2021jvn}.}.

\begin{figure}[!htb]
\centering
\includegraphics[width=0.9\columnwidth, angle=0]{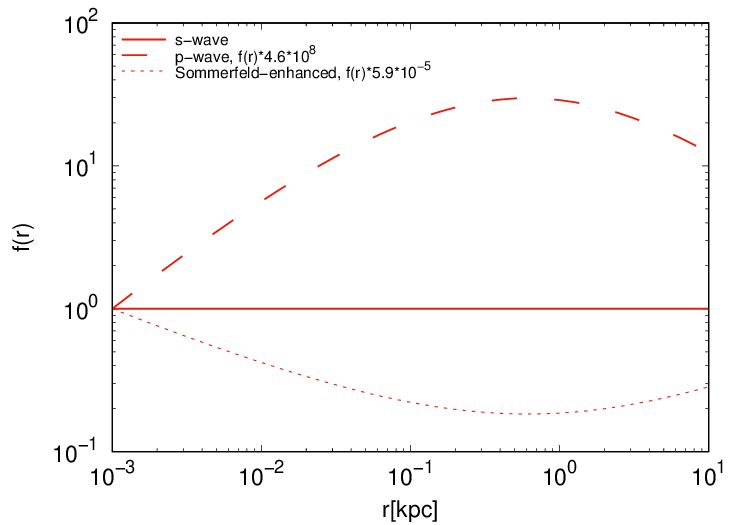}
\caption{The $f(r)$ of the DM density profile with $\rho_s=3\times10^8 \rm M_{\odot}/\rm kpc^3 $ and $r_s =0.8 \rm kpc$ for s-wave (solid line), p-wave (dashed line) and Sommerfeld-enhanced (dotted line) scenarios. Note that all lines have been normalized to the same values at $10^{-3} \rm kpc$.}
\label{fig:fr}
\end{figure}

The averaged $\left<\sigma v\right>$ at any position is given by
\begin{eqnarray}\label{eq:zy10}
\left<\sigma v\right> = a \int\int F(v_{\mathrm{rel}}) f(v_1, \textbf{r})f(v_2, \textbf{r}) \mathrm{d}v_1^3 \mathrm{d}v_2^3,
\end{eqnarray}
where $f(v, \textbf{r})$ represents the DM velocity profile at position $\textbf{r}$.
We assume that the DM velocity distribution follows a standard isotropic Maxwell-Boltzmann distribution~\cite{Fairbairn:2008gz,Vogelsberger:2008qb,Zemp:2008gw,Kuhlen:2009vh}.
The DM velocity dispersion at $\textbf{r}$ is determined by the Jeans equation for DM, from which the radial velocity dispersion $\sigma_{D,r}^2$ of DM can be solved. These equations have the same forms as Eq. \ref{eq:zy1} and Eq. \ref{eq:zy2}, except for the density, radial velocity dispersion, and velocity anisotropy replaced by those of DM.
Due to insufficient information on the DM velocity dispersion in dwarf galaxies, we assume that the velocity dispersion anisotropy of DM is zero.
In the non-relativistic limit, $\left<\sigma v\right>$  becomes a function of $r$, which is given by
\begin{eqnarray}\label{eq:zy11}
\left<\sigma v\right> &=& a \int \sqrt{\frac{2}{\pi}} \frac{1}{v_{p}^3} v_{\mathrm{rel}}^2 e^{-\dfrac{v_{\mathrm{rel}}^2}{2v_{p}^2}}F (v_{\mathrm{rel}})\mathrm{d}v_{\mathrm{rel}} \nonumber \\
&\equiv & a\cdot f(r) ,
\end{eqnarray}
where $v_p^2 \sim 2 \sigma_{D,r}^2$. It is evident that $f(r)=1$ corresponds to s-wave annihilation. For illustrative purposes, we present some examples of  $f(r)$ in Fig. \ref{fig:fr}.

We define the effective J-factor as
\begin{eqnarray}\label{eq:zy12}
J &=& \int_{\Delta\Omega} \int_{l.o.s.} \rho^{2}(r) f(r) \mathrm{d}l\mathrm{d}\Omega
\nonumber\\
&=&\int_0^{\theta_\mathrm{max}}\int_{\theta\cdot d}^{r_\mathrm{max}}\frac{4\pi r\sin\theta\rho^{2}(r)}{\sqrt{r^2-(\theta\cdot d)^2}}f(r)\mathrm{d}r\mathrm{d}\theta,
\end{eqnarray}
where $\theta_\mathrm{max}$ is the maximum integral angle, $r_\mathrm{max}$ is the maximum radius of the dSph, and $d$ is the heliocentric distance of the dSph.
We calculate the J-factor and effective J-factors within an integral angle of $0.5^\circ$ for each obtained density profile. The statistical results derived from all derived density profiles for the s-wave, p-wave, and Sommerfeld-enhanced annihilation scenarios are shown in Tab. \ref{tab:jfac}. It is evident that the J-factor and effective J-factors of Ursa Major III are larger than those of other dSphs, as summarized in  Ref.~\cite{Boddy:2019qak}.

\begin{table}[!htb]
\belowcaptionskip=1em
\centering
\caption{The J-factor and effective J-factors with integral angle $0.5^\circ$ for Ursa Major III.}
\setlength\tabcolsep{1.0em}
 \begin{tabular}{lcc}
  \hline\hline\noalign{\smallskip}
   \multirow{2}{*}{n}      & $\log_{10}J$  \\
                              &$(\log_{10}[\GeV^{2}\cm^{-5}])$   \\
  \hline\noalign{\smallskip}
  0(s-wave)                    & $21.4_{-0.7}^{+0.7}$   \\
  2(p-wave)                    & $13.6_{-1.6}^{+1.4}$   \\
  -1(Sommerfeld-enhanced)      & $25.3_{-0.5}^{+0.5}$   \\
  \noalign{\smallskip}\hline\hline
\end{tabular}
\label{tab:jfac}
\end{table}

The expression for the $\gamma$-ray flux from DM annihilation in an energy bin can be expressed as
\begin{eqnarray}\label{eq:zy13}
\Phi &=& \frac{1}{8\pi m_{DM}^2}\int \frac{\mathrm{d}N_\gamma}{\mathrm{d}E_\gamma}\mathrm{d}E_\gamma\int_{\Delta\Omega} \int_{l.o.s.} \left<\sigma v\right> \rho^{2} \mathrm{d}l\mathrm{d}\Omega
\nonumber \\
&=& \frac{a\cdot J}{8\pi m_{DM}^2}\int \frac{\mathrm{d}N_\gamma}{\mathrm{d}E_\gamma}\mathrm{d}E_\gamma,
\end{eqnarray}
where $m_{DM}$ is the DM mass, and $\frac{\mathrm{d}N_\gamma}{\mathrm{d}E_\gamma}$ is the differential $\gamma$-ray spectrum from an annihilation event, which has been provided by PPPC4DM ~\cite{Cirelli:2010xx,Ciafaloni:2010ti}.
The term $(v_{\mathrm{rel}}/c)^n$ in Eq. \ref{eq:zy9} is already included in the factor $J$ in Eq. \ref{eq:zy13}.
If $n=0$, then this factor corresponds to the commonly used velocity-independent J-factor, with the constant $a$ corresponding to $\left<\sigma v\right>$.
For $n \neq 0$, such as $n=2$ or $n=-1$ as discussed in this study, this factor represents the velocity-dependent effective J-factor.
In this scenario, the value of local $\left<\sigma v\right>$ in the Milky Way also depends on the local velocity dispersion, which is typically $\sim 270~\km/\s$.

The absence of a $\gamma$-ray excess from Ursa Major III in 15 years of Fermi-LAT data has been reported in both the point-source and the extended-source analyses~\cite{Crnogorcevic:2023ijs}.
The authors of Ref.~\cite{Crnogorcevic:2023ijs} calculated the upper limits on the signature flux by assuming a point source at the location of Ursa Major III, and presented the corresponding likelihood profile, which illustrates the log-likelihood for $\gamma$-ray signature fluxes in the energy range between 500 MeV and 500 GeV.
Using this profile, the upper limits on the DM annihilation cross section from Ursa Major III can be determined.
In setting the upper limits on the DM annihilation cross section, we incorporate the uncertainty of the J-factor in the likelihood, which is given by
\begin{eqnarray}\label{eq:zy14}
\mathcal{L}&&=\prod_{i}\mathcal{L}_{i} (\Phi_i)\nonumber\\
&&\times \frac{e^{-[\log_{10}(J)-\log_{10}(J_\mathrm{med})]^{2}/2\sigma^{2}}}{\ln(10)J_\mathrm{med}\sqrt{2\pi}\sigma},
\end{eqnarray}
where $\Phi_i$ is the expected signature flux in the $i$-th energy bin for $\Phi_i$, $\mathcal{L}_i$ is the corresponding likelihood, and $J_\mathrm{med}$ and $\sigma$ are the median value and deviation of the J-factor, respectively.
The upper limits on the $\left<\sigma v\right>$ at $95\%$ confidence level are determined by requiring that the corresponding $\log\mathcal{L}$ value decreases by $2.71/2$ from its maximum value.

\begin{figure*}[!htb]
\centering
\includegraphics[width=0.9\columnwidth, angle=0]{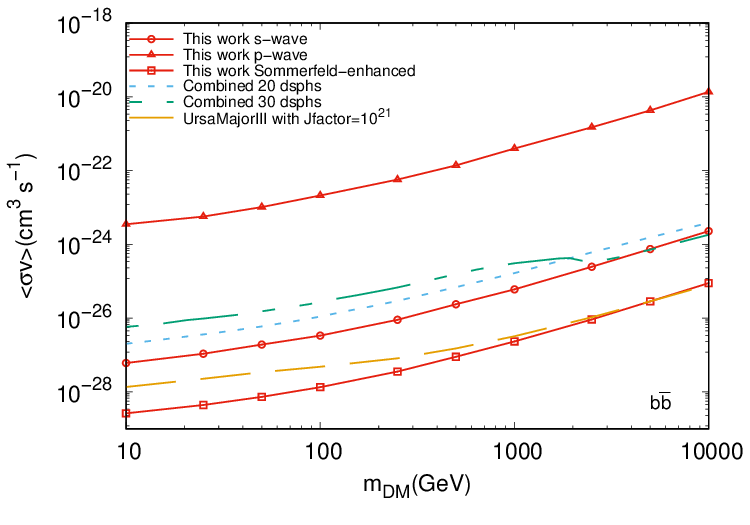}
\includegraphics[width=0.9\columnwidth, angle=0]{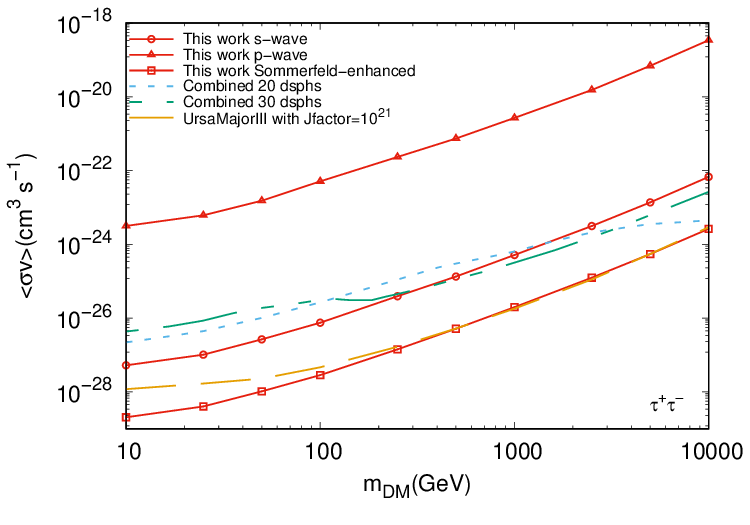}
\caption{The constraints on local DM annihilation cross sections at $95\%$ C.L. for the $b\bar{b}$ (left)
and $\tau^+\tau^-$ (right) annihilation channels. The red solid lines marked with circles, triangles, and squares represent the results for s-wave($n=0$), p-wave ($n=2$) scenario, and Sommerfeld-enhanced ($n=-1$) annihilations, respectively. The dotted, dash-dot-dotted, dashed lines represent the constraints derived from the combination of 20 dSph observations performed by Fermi-LAT, HAWC, H.E.S.S., MAGIC, and VERITAS collaborations~\cite{Hess:2021cdp}, the 30 dSph observations by Fermi-LAT~\cite{McDaniel:2023bju}, and the Ursa Major III observation by Fermi-LAT with a J-factor of $J=10^{21}\GeV^{2}\cm^{-5}$~\cite{Crnogorcevic:2023ijs}, respectively.
}
\label{fig:zy3ab}
\end{figure*}

The upper limits on the local $\left<\sigma v\right>$ in the Milky Way for the $b\bar{b}$ and $\tau^+\tau^-$ annihilation channels are shown in Fig. \ref{fig:zy3ab}. 
The red lines, marked with circles, triangles, and squares, represent the results for s-wave  ($n=0$), p-wave  ($n=2$), and Sommerfeld-enhanced ($n=-1$) scenarios, respectively.
Notably, the constraints on
$\left<\sigma v\right>$ from Ursa Major III 
are remarkably stringent for velocity-independent annihilation. These constraints directly preclude  the thermal relic annihilation cross section for DM masses below the TeV scale for the $b\bar{b}$ channel. This is primarily attributed to the dwarf galaxy's relatively large J-factor and its proximity to Earth. 

For comparison, the constraints derived from the combination of 20 dSph observations performed by Fermi-LAT, HAWC, H.E.S.S., MAGIC, and VERITAS collaborations~\cite{Hess:2021cdp}, the 30 dSph observations by Fermi-LAT~\cite{McDaniel:2023bju}, and the Ursa Major III observation by Fermi-LAT with a J-factor of $J=10^{21}\GeV^{2}\cm^{-5}$~\cite{Crnogorcevic:2023ijs} are also shown in Fig. \ref{fig:zy3ab}.
In comparison to the combined results from 20 dSphs and 30 dSphs, our s-wave constraints exhibit stringency for DM masses below $\mathcal{O}(1)$ TeV. For the $\tau^+ \tau^-$ channel, the constraints from 20 dSphs provided by Ref.~\cite{Hess:2021cdp} are stricter above several TeVs,  owing to the dominant constraints set by ground-based detectors in this mass region. Note that the Ursa Major III constraints of \cite{Crnogorcevic:2023ijs} shown in Fig. \ref{fig:zy3ab} are established with a fixed J-factor. Although the J-factor of $J=10^{21}\GeV^{2}\cm^{-5}$ for those constraints closely aligns with the median value of our J-factor, our constraints include the uncertainties associated with the J factor. Consequently, our constraints are less stringent than those particular results.

In the p-wave annihilation scenario, the constraints on the local annihilation cross section can be relaxed by three orders of magnitude, while in the Sommerfeld-enhanced scenario, the constraints can be strengthened by one order of magnitude.
For p-wave annihilation, the constraints of this study are comparable with those given by \cite{Zhao:2017dln}, which are derived from three sizable ultrafaint dSphs. The three dSphs considered in Ref.~\cite{Zhao:2017dln} encompass Willma 1, Reticulum II, and Triangulum II, characterized by effective J-factors of $\log_{10}J=11\pm 2.3$, $10.6\pm2.7$, and $14.4\pm2.3\GeV^{2}\cm^{-5}$ \footnote{The effective J-factor defined in this study is related to the $C$ factor defined in Ref.~\cite{Zhao:2017dln} via $J = C/c^2$ for p-wave annihilation and $J = C \times c$ for
Sommerfeld-enhanced annihilation, where $c$ represents the speed of light. These effective J-factors are consistent with those derived by other studies, that are summarized in Ref.~\cite{Boddy:2019qak}.}, respectively. Despite Triangulum II having a substantial median effective J-factor value, the uncertainties in its kinematic observations yield a large deviation in the J-factor, thereby resulting in comparable constraints when compared to Ursa Major III.

\section{Impact of Excluding the Largest Velocity Outlier}
\label{Sec.IV}

The preceding analysis is based on the assumption of 11 radial velocity members of Ursa Major III. However, by excluding the largest velocity outlier, which has the lowest radial velocity among the members, a marked reduction in velocity dispersion is observed. Furthermore, removing an additional outlier results in an unresolved dispersion. The uncertainty of this measurement would lead to substantial variations in the J-factor. In this section, we exclude the largest velocity outlier and derive the DM profiles based on the residual 10 members.

\begin{figure}[!htb]
\centering
\includegraphics[width=0.9\columnwidth, angle=0]{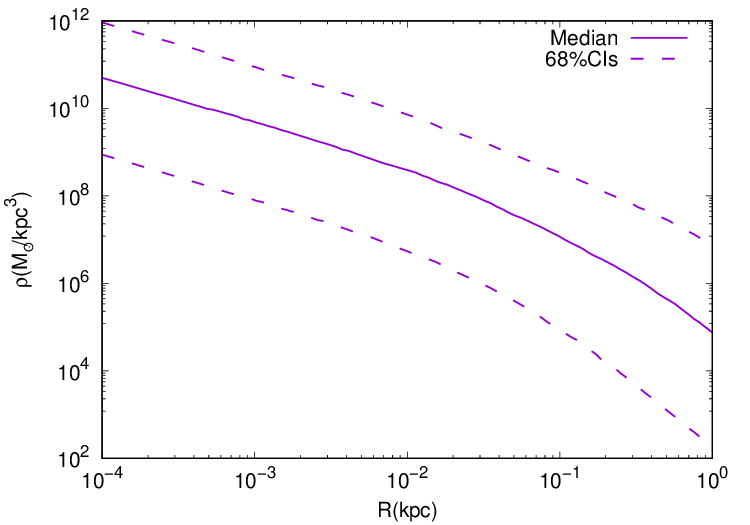}
\caption{Similar to Fig. \ref{fig:zy2}, but for Ursa Major III with 10 member stars, excluding the largest velocity outlier.}
\label{fig:zy4}
\end{figure}

For this analysis, we generate eight MCMC chains, each consisting of $10^5$ iterations. The DM density of Ursa Major III at varying distances from the galactic center, as determined from 10 member stars, is shown in Fig. \ref{fig:zy4}.
In comparison to Fig. \ref{fig:zy2}, which incorporates data from 11 member stars, the median value in Fig. \ref{fig:zy4} is lower by one order of magnitude.
The uncertainties for the DM densities at the $68\%$ CIs in this figure are noticeably expanded, with the lower limits being particularly pronounced.
As reported in Ref. ~\cite{smith2023discovery}, excluding the star with the largest velocity outlier diminishes the intrinsic velocity dispersion from $3.7^{+1.4}_{-1.0} \km/\s$ to $1.9^{+1.4}_{-1.1} \km/\s$.
This suggests a large uncertainty in the intrinsic velocity dispersion, leading to a wide range of DM density uncertainties.

Excluding the largest velocity outlier, we conduct a Jeans analysis for 10 member stars, and obtain $\log_{10}J = 17.7_{-3.9}^{+2.5} \GeV^{2}\cm^{-5}$ for the s-wave velocity-independent scenario. This median value is notably  lower than that obtained for 11 member stars, and the associated deviation is significant larger. The J-factor calculated in this analysis is notably smaller than the range of $\sim 10^{19}-10^{22} \GeV^{2}\cm^{-5}$ provided by Ref.~\cite{Errani:2023sgd}, which is based on the intrinsic velocity dispersion and the analytic formula. The intrinsic velocity dispersion assumes all $\sigma^2_p(R_i)$ to be uniform in Eq.~\ref{eq:zy4}, neglecting the positional information of member stars in the velocity distribution. This method differs from the Jeans analysis conducted in this study, resulting in distinct J-factors.

It is noteworthy that an excessively small J-factor implies a markedly low DM density in the galaxy. As discussed in Ref.~\cite{Errani:2023sgd}, the diminished DM density at the core of Ursa Major III poses a challenge in mitigating the tidal effects. For a rough assessment, we use the Roche criterion to determine the tidal radius $r_t$ \cite{Evans:2003sc}, beyond which the DM density is expected to undergo a substantial reduction due to the tidal effects:
\begin{eqnarray}\label{eq:zy15}
\frac{M_{\rm dSph}(r_t)}{r_t^3} = \frac{M_{\rm MW}(r_{\rm dSph}-r_t)}{(r_{\rm dSph}-r_t)^3},
\end{eqnarray}
where $M_{\rm dSph}(r)$ and $M_{\rm MW}(r)$ denote the enclosed masses of the dSph and Milky Way within the radius of $r$, respectively, and $r_{\rm dSph}$ is the distance between their centers. We find that a portion of the derived density profiles of Ursa Major has an angle radius $\theta_t \equiv r_t/r_{\rm dSph}$ less than $0.5^\circ$. Considering an integration angle of $\min(\theta_t, 0.5^\circ)$, we derive the J factor as $\log_{10}J = 17.7_{-4.3}^{+2.5} \GeV^{2}\cm^{-5}$, which encompasses a broader $1\sigma$ boundary. If we disregard  density profiles with a small angle radius $\theta_t< 0.5^\circ$, we derive a higher J factor as $\log_{10}J = 19.2_{-1.8}^{+1.6} \GeV^{2}\cm^{-5}$.

In the analysis involving 10 member stars, the effective J-factors for p-wave annihilation and Sommerfeld-enhanced annihilation are calculated as $\log_{10}J = 8.1_{-6.1}^{+4.0} \GeV^{2}\cm^{-5}$ and $ 22.7_{-2.8}^{+1.9} \GeV^{2}\cm^{-5}$, respectively. If we impose a requirement that the density profiles possess an angular radius $\theta_t\geq 0.5^\circ$, we can derive effective J factors as $10.2_{-2.6}^{+2.6} \GeV^{2}\cm^{-5}$ for p-wave annihilation and $23.8_{-1.4}^{+1.1} \GeV^{2}\cm^{-5}$ for Sommerfeld-enhanced annihilation, respectively.

These J-factors result in constraints on the DM annihilation cross section that are significantly weaker than those previously obtained  from data involving 11 member stars by several orders of magnitude, making these constraints less impactful. It is crucial to emphasize that the numerical results of the analysis involving 10 member stars in this section are derived from a highly limited set of kinematic data, and warrant further scrutiny through future observations.  Furthermore, the tidal effect necessitates further exploration in subsequent studies.

\section{Conclusion}

The dSph Ursa Major III emerges as a particularly promising candidate for detecting the $\gamma$-ray signatures from DM annihilation. This distinction is attributed to its close proximity and the potential for a high J-factor, pending confirmation through further measurements. The composition of DM within this galaxy is significantly influenced by the precise identification of member stars. Assuming the inclusion of all 11 observed member stars within Ursa Major III, we employ Jeans analysis to extract the DM density profiles of this galaxy. A series of profiles is derived based on the posterior probability derived from fitting the kinematic data.

Utilizing the derived DM density profiles, we calculate the J-factor for s-wave annihilation, along with the effective J-factors for p-wave and Sommerfeld-enhanced annihilation scenarios.
Employing the likelihood map of Ursa Major III, we establish constraints on the $\gamma$-ray flux from DM annihilation. For the $b\bar{b}$ and $\tau^+\tau^-$ annihilation channels, we set the upper limits on DM annihilation cross sections at $95\%$ C.L.. Notably, this dSph provides extremely stringent constraints, surpassing those derived from the joint analysis of numerous previously studied dwarf galaxies. Exploring the velocity-dependent annihilation scenarios yields valuable insights for interpreting results from DM indirect detection experiments. These scenarios have the potential to introduce constraints that diverge from the conventional limits established through dSph $\gamma$-ray observations.

The above results highly depend on the precise determination of the J-factor, with the identification of individual stars within Ursa Major III playing a pivotal role in shaping the final J-factor estimate. The exclusion of a single member star exhibiting the largest velocity outlier can lead to a significant reduction in velocity dispersion, consequently causing a notable increase in the uncertainty of the J-factor.
In the absence of this outlier, the significantly low J-factors, compounded by large uncertainties, substantially relax the constraints on the DM annihilation cross section. 
Future, more comprehensive observations of the member stars comprising this dwarf galaxy hold promise for refining our understanding of its J-factor. Such detailed observations will play an important role in enhancing the precision of the J-factor determination.

\acknowledgments{This work is supported by the National Natural Science Foundation of China under Grants Nos. 11947005, 12175248, 12205388, 12227804, the Science\&Technology Development Fund of Tianjin Education Commission for Higher Education No. 2020KJ003, and the PhD research startup foundation of Tianjin Normal University under Grant No.52XB1912.
}



\bibliography{ursamajorIII}

\end{document}